\documentclass[%
reprint,
superscriptaddress,
 amsmath,amssymb,
aps,
pra,
]{revtex4-1}

\pdfoutput=1
\usepackage{journal}
\usepackage{graphicx}
\usepackage{dcolumn}
\usepackage{bm}

\begin{document}


\title{Dispersion and light transport characteristics of large-scale photonic-crystal coupled nanocavity arrays}

\author{Nobuyuki Matsuda}
\email{m.nobuyuki@lab.ntt.co.jp}
\affiliation{NTT Nanophotonics Center, NTT Corporation, 3-1 Morinosato-Wakamiya, Atsugi, Kanagawa 243-0198, Japan}
\affiliation{NTT Basic Research Laboratories, NTT Corporation, 3-1 Morinosato-Wakamiya, Atsugi, Kanagawa 243-0198, Japan}

\author{Eiichi Kuramochi}
\affiliation{NTT Nanophotonics Center, NTT Corporation, 3-1 Morinosato-Wakamiya, Atsugi, Kanagawa 243-0198, Japan}
\affiliation{NTT Basic Research Laboratories, NTT Corporation, 3-1 Morinosato-Wakamiya, Atsugi, Kanagawa 243-0198, Japan}

\author{Hiroki Takesue}
\affiliation{NTT Basic Research Laboratories, NTT Corporation, 3-1 Morinosato-Wakamiya, Atsugi, Kanagawa 243-0198, Japan}

\author{Masaya Notomi}
\affiliation{NTT Nanophotonics Center, NTT Corporation, 3-1 Morinosato-Wakamiya, Atsugi, Kanagawa 243-0198, Japan}
\affiliation{NTT Basic Research Laboratories, NTT Corporation, 3-1 Morinosato-Wakamiya, Atsugi, Kanagawa 243-0198, Japan}


\begin{abstract}
We investigate the dispersion and transmission property of slow-light coupled-resonator optical waveguides that consist of more than 100 ultrahigh-$Q$ photonic crystal cavities. We show that experimental group-delay spectra exhibited good agreement with numerically calculated dispersions obtained with the three-dimensional plane wave expansion method. Furthermore, a statistical analysis of the transmission property indicated that fabrication fluctuations in individual cavities are less relevant than in the localized regime. These behaviors are observed for a chain of up to 400 cavities.
\end{abstract}

\pacs{Valid PACS appear here}
\maketitle


Slow light on a chip \cite{1,2} provides many potential applications including ultrasmall optical buffers \cite{3,4}, passive photonic components with enhanced phase sensitivity \cite{5} and all-optical signal processing with enhanced light-matter interactions \cite{6}. A coupled-resonator optical waveguide (CROW), which is a one-dimensional array of nearest-neighbor-coupled identical optical cavities, is an attractive candidate for slow light waveguides \cite{7,8,9}. The guided mode in a CROW is a collectively resonant supermode of the constituent cavities, whose dispersion property can be greatly modulated simply by changing the inter-cavity coupling strengths. Using CROW, we can easily obtain a small group velocity, {\it i.e.}, a large group index $n_{\rm g}$, in a transmission band that is much wider than that of the constituent cavities.

It is important to implement a large-scale CROW for certain applications including optical buffers. To achieve a meaningful transmission of light in a large-scale resonator chain, both a high cavity $Q$ and a small resonator footprint are important, as well as for realizing a large $n_{\rm g}$ \cite{,4}. Chains consisting of more than a hundred resonators have already been developed using short waveguide elements \cite{10}, microring resonators \cite{3,11}, microdisk resonators \cite{12}, and photonic crystal (PC) cavities \cite{4,13}. Among them, the PC cavity is of particular interest, because it accommodates both a cavity $Q$ as high as one million and a mode volume as small as a wavelength. Indeed, optical pulse propagation in 400 coupled cavities \cite{13} and an $n_{\rm g}$ as large as 170 \cite{4} were demonstrated using PC cavities. These are record values for any CROW. Moreover, the CROW has proven useful for buffering non-classical states of light with a certain transit time tunability \cite{13}.

The guided mode of a CROW can be approximated as an optical analogue of the tight-binding model, in which non-nearest-neighbor cavity couplings are assumed to be negligible \cite{8}. This is, however, an approximated situation and our PC CROWs exhibited small deviations from the model \cite{4}. Thus, it is important to compare the experiment with the result of a rigorous dispersion calculation for a better characterization of the device. In addition, a practical device is accompanied by some degree of fabrication fluctuation, which results in the fluctuation of each $Q$ and the inter-cavity coupling strength. Light interference caused by the fluctuations is responsible for a localized mode, which degrades the transmission property and the maximum available $n_{\rm g}$ in the large-scale regime \cite{10}. To investigate the localization effect, Cooper {\it et al.} performed a statistical analysis of the transmission properties, showing the non-localized nature of light transport in a microring resonator-based CROW up to 235 resonators \cite{14}. Such information gives us the degree of structural uniformity, which is an important specification of the accuracy of the current fabrication technology.

In this paper, we theoretically and experimentally investigate the dispersion and transmission properties of CROWs consisting of several hundred high-$Q$ PCs. First, we rigorously calculate the photonic band structure using a three-dimensional vector plane wave expansion (PWE) method, and discuss the discrepancy with the tight-binding model. Next, we show that the numerical results agree well with the experimental dispersions of the CROWs obtained with a pulsed time-of-flight (TOF) method. We also statistically investigate the light-transport property of the CROW to show that our CROW is fabricated in a manner less relevant than in the localized regime.

We consider CROWs fabricated on an air-bridged 2-D Si PC slab with a triangular lattice of air holes (Fig. 1). The lattice constant $a$ is 420 nm and the slab thickness $t$ is 0.5$a$. Each cavity is formed by the local width modulation of a barrier line defect with a width of 0.98$a \sqrt{3}$ (W0.98). The red and green holes are shifted by 8 and 4 nm, respectively, along transverse directions. This yields a cavity $Q$ as high as $10^6$ and a cavity mode volume of 1.7 $(\lambda/n)^3$ \cite{4, 15}. We study a CROW with a cavity pitch $L_{\rm cc}$ of 5$a$.

\begin{figure}
	\centering\includegraphics[width=8.6cm]{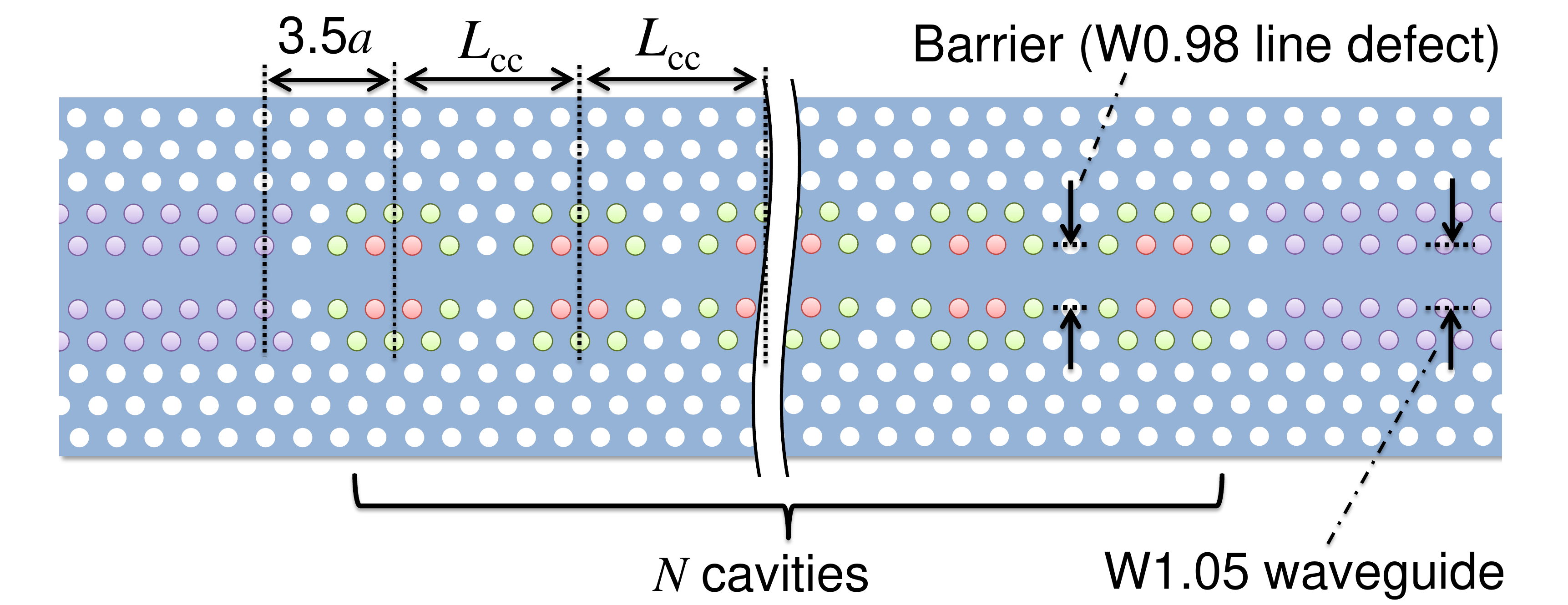}
	\caption{A CROW consisting of photonic-crystal mode-gap nanocavities for a cavity pitch $L_{\rm cc} = 5a$.}
\end{figure}

We calculate the band structure of the CROW with a three-dimentional vector PWE method using BandSOLVE (Synopsys, Inc.). A computational supercell is shown in Fig. 2(a). The refractive index of the Si slab is set at 3.47. We start from a W0.98 line defect waveguide without any shifting of the holes. The calculated dispersion diagram is shown in Fig. 2(b) (black filled squares). Here the wavenumber is the component along the $z$ CROW axis ({\it i.e.}, $\Gamma$-K direction). Here the band folding appears as a result of the artificial periodicity of the 5$a$ super cell imposed to the waveguide. Next, with the hole shifts (red filled circles), we see the bands anti-cross at the folding point at a wavenumber of 0, and a mini gap appears due to the physical periodicity imposed by the cavities. Then, the lower band is isolated across the band edge of the W0.98 barrier, which is at around a normalized frequency of 0.2724 (indicated by the horizontal blue line). Since the resonance mode of a single mode-gap cavity is formed just below the band edge of the barrier line defect, the isolated lower band is considered to be the coupled-cavity mode. We also see that the isolated lower band has flat dispersions on both ends. This resembles a cosine-shaped dispersion, which is a CROW dispersion obtained under the tight-binding approximation \cite{8}. Similar formation mechanisms can also be seen in CROWs based on other PC cavity designs \cite{16, 17}. Note that in our case the entire redshift of the CROW bands is considered to originate from an increase in the effective line-defect width imposed by the hole shifts.

Figure 2(c) shows $n_{\rm g}$ spectra for the corresponding range of frequency of Fig. 2(b), where $n_{\rm g} = c/v_{\rm g}$, $c$ is the speed of light in a vacuum and $v_{\rm g}$ is the group velocity calculated by taking a derivative of the dispersion diagram shown in Fig. 2(b). The line-defect mode exhibits an $n_{\rm g}$ spectrum that increases monotonically as the frequency decreases, {\it i.e.}, an increase in wavelength as has already been shown experimentally \cite{18}. Then, the existence of the mini gap of the CROW raises the $n_{\rm g}$ spectrum toward a larger $n_{\rm g}$ at a mini-gap frequency of around 0.273. As a result, we can obtain an isolated CROW band below the mini gap, with a zero-dispersion wavelength in the large $n_{\rm g}$ regime.

\begin{figure}
	\centering\includegraphics[width=8.6cm]{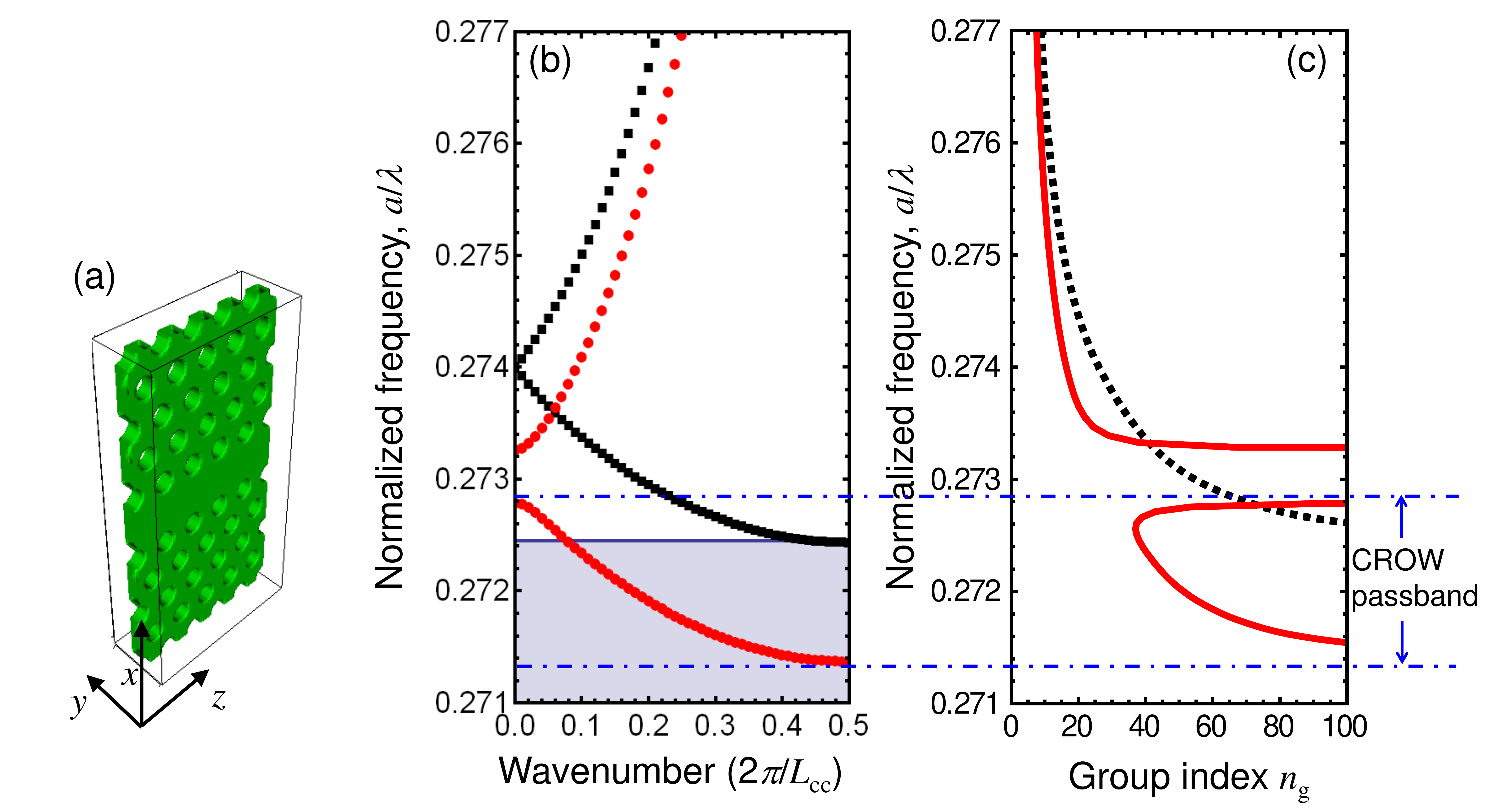}
	\caption{A supercell used for the PWE calculation. Dimensions: $10a \times 2a \times 5a$. (b) The simulated dispersion relation with (black circles) and without (red squares) the nanocavity, {\it i.e.}, the hole shifts. The blue-hatched region indicates the mode gap of the W0.98 line-defect waveguide. (c) Calculated group index spectra for the CROW (red solid curve) and W0.98 waveguide (black dashed curve) with a hole radius of 110 nm. }
\end{figure}

The $n_{\rm g}$ spectrum of our CROW represents an asymmetric shape, distinct from the symmetric $n_{\rm g}$ spectrum with respect to the center frequency obtained from cosine-like dispersion for the tight-binding model. This is because the cavities of our CROW are based on the width-modulation of the line defect mode. From the tight-binding picture, the deviation would presumably be associated with finite non-nearest neighbor couplings.

We fabricated CROW samples with cavity numbers $N$ of 200 and 400. Access waveguides were fabricated to provide a connection between the CROW and external tapered optical fibers for the coupling. (Note that the $N$ = 400 sample is the same as that used in \cite{13}.) The access waveguides were W1.05 line-defect PC waveguides (the region of purple holes in Fig. 1) and Si wire waveguides (not shown). The cavity pitch was apodized at the waveguide-cavity connection as shown in Fig. 1. We also fabricated reference waveguides, in which the CROW section was replaced by a W1.05 line defect waveguide. The overall PC section was 860 $\mu$m long for all the samples. Figure 3 shows the linear transmission spectra of the CROWs and the reference waveguide. Each CROW exhibits a clear passband with isolations over 30 dB. We can also see a decrease in the transmission bandwidth for a large $N$. This is because larger $n_{\rm g}$ modes suffer larger propagation losses. For a detailed discussion of bandwidth reduction including the effect of the coupling between the access waveguide and the CROW see \cite{4}.

We investigate the group-delay dispersion of the CROWs by using the pulsed TOF method. We performed this direct time-domain measurement because it is difficult to infer the dispersion from the transmission spectrum in accordance with the procedure in \cite{4} for large-scale CROWs, whose passband is quasi-continuous and where the spectral peaks of the extended mode are no longer distinguishable. Such a direct transmission measurement is also important to show the performance of the CROW as a transmission line, rather than an indirect measurement using, for example, out-of-plane radiations \cite{17, 19}.

\begin{figure}
	\centering\includegraphics[width=8.6cm]{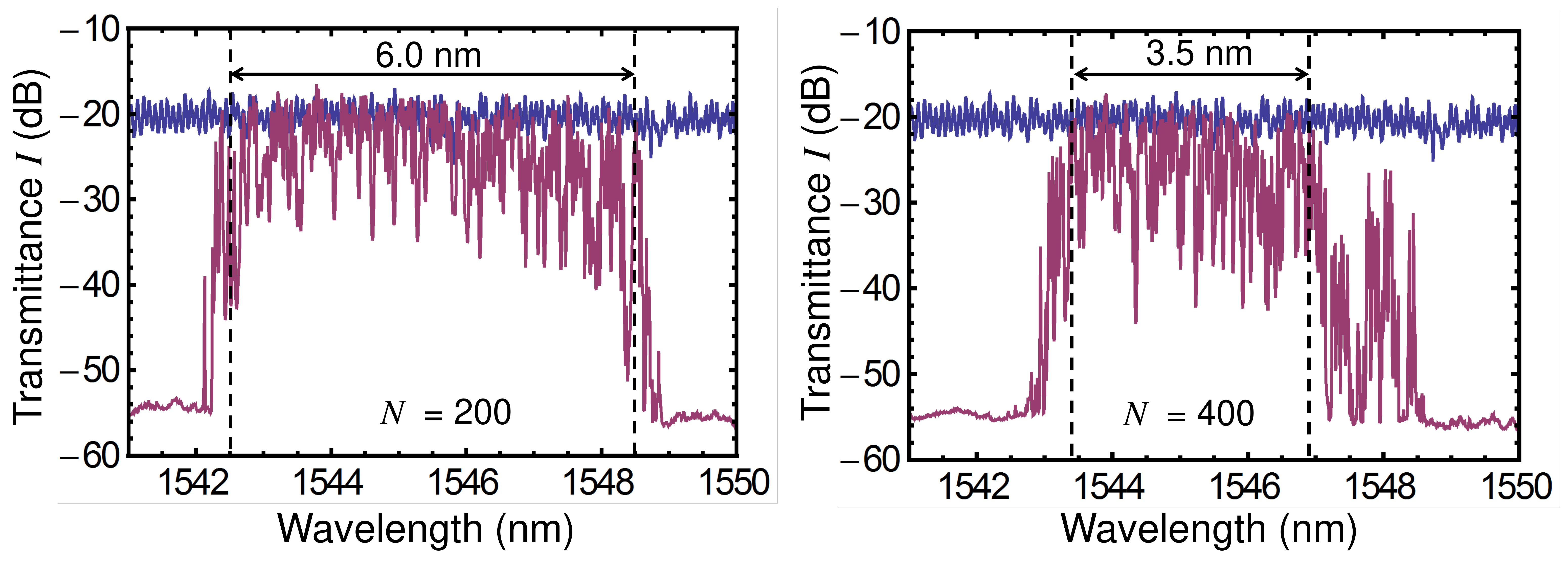}
	\caption{Transmission spectra of CROWs (red curves). The blue curve shows the data for the reference W1.05 waveguide. Using the CROW data, we perform a statistical analysis of the light transport in each wavelength range surrounded by vertical dashed lines.}
\end{figure}

For the TOF measurement, input optical pulses (duration: 80 ps) were obtained from a wavelength-tunable cw laser followed by an intensity modulator. Output optical pulses from the samples were directly detected with a high-speed optical sampling oscilloscope (bandwidth: 80 GHz) synchronized with a gate signal applied to the intensity modulator. Fig. 4(a) is a density plot of the output-waveform spectra obtained for a reference waveguide and the CROWs. The horizontal axis shows the center frequency of each optical pulse, while the vertical axis is the time relative to the arrival of the gate signal. The black dots indicate the temporal positions of pulse peaks at each wavelength. We can see clear propagation delays for the CROWs up to $N$ = 400 entirely in the band. Note that the peak temporal positions were plotted only for the waveform data that exhibited peak intensities greater than five standard deviations of the background level.

We show $n_{\rm g}$ spectra in Fig. 4(b). The data are extracted from the difference between the traveling times (dotted data points) of the reference waveguide and a CROW with the length of each CROW taken into account. Here, the $n_{\rm g}$ value of the reference waveguide is assumed to be 6, which is obtained from a simulation. From the data, we find that $n_{\rm g}$ is constant regardless of $N$. Hence, as expected the dispersion in our CROW is determined solely by local coupling structures \cite{4}.

The solid curves in Fig. 4(b) are the simulation results shown in Fig. 2(c). We see the numerical result shows good agreement with the experimental data, including the asymmetric shape of the $n_{\rm g}$ spectra for the CROW band. Some deviations from the experimental data is because $n_{\rm g}$ is sensitive to the small change in the hole radius $r$ in our CROW. For example, the dot-dashed and dashed curves represent the numerical dispersions when $r$ differs only by $\pm$ 2 nm. Such a small variation is in the range of the fabrication error. Note that we arbitrarily shifted the frequency of the numerical dispersion to fit the wavelength to that of experimental data. This operation corresponds to imposing a variation in $t$, by which the shape of $n_{\rm g}$ spectra remain unchanged.

\begin{figure}
	\centering\includegraphics[width=8.6cm]{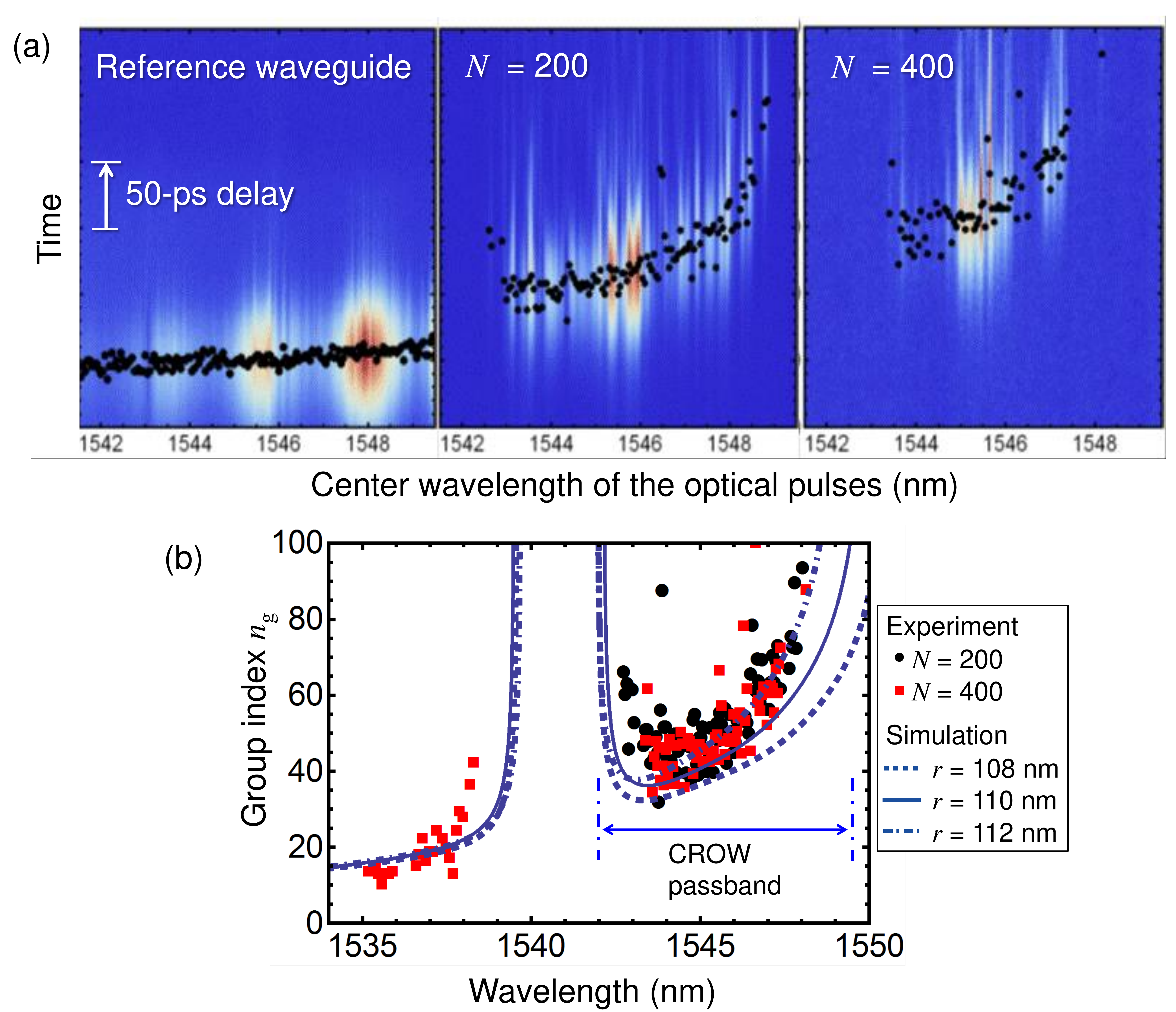}
	\caption{(a) Spectra of output waveforms for the CROWs and the W1.05 reference waveguide. The black dots are the peak temporal positions of each waveform. (b) Extracted group index spectra for various samples. The curves show the results of a numerical simulation with the PWE method for different hole radius $r$.}
\end{figure}

\begin{figure}
	\centering\includegraphics[width=6cm]{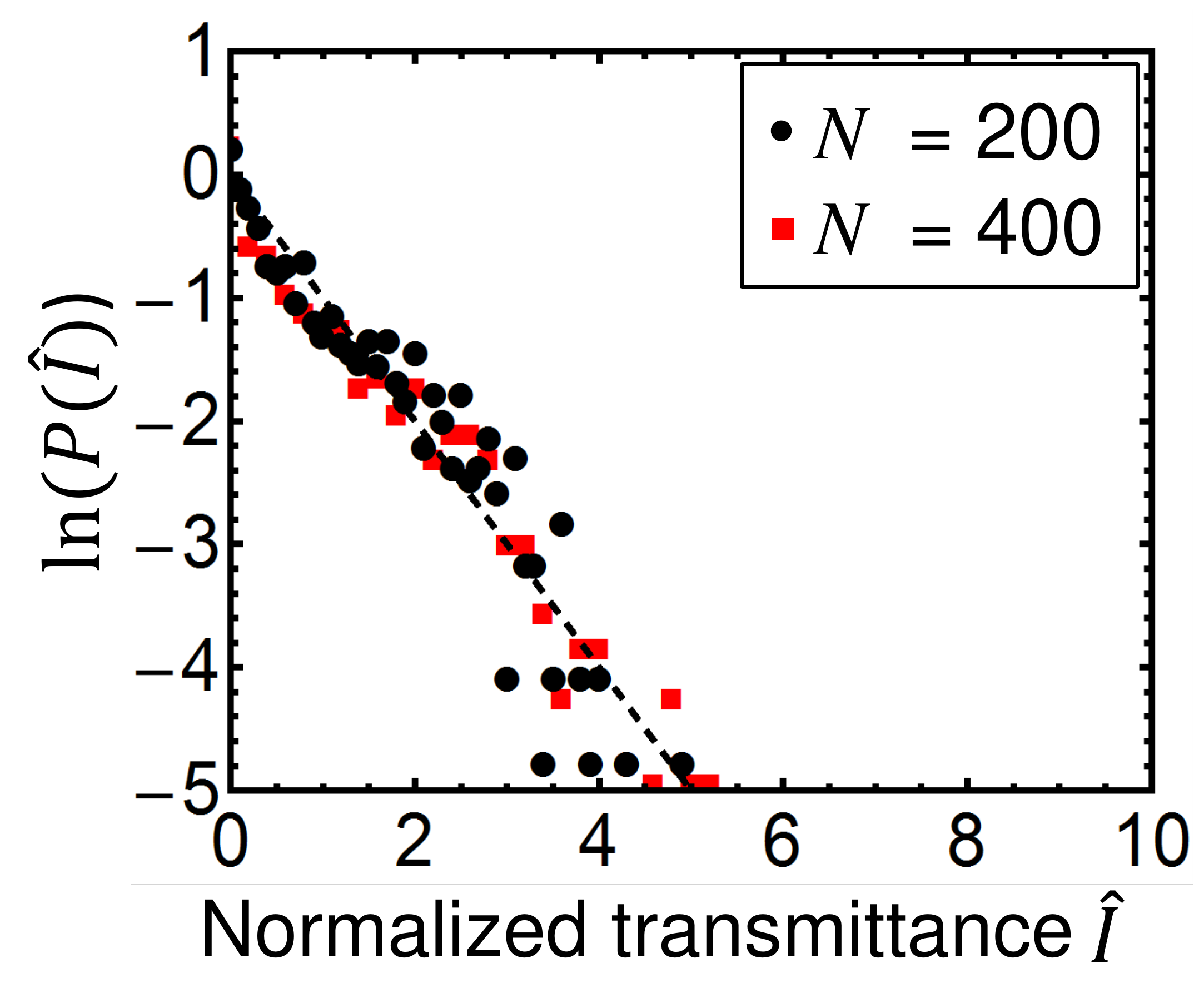}
	\caption{The probability distribution functions of the normalized intensity transmittance for wavelength ranges indicated by arrows in each panel of Fig. 3. The dashed line shows the Rayleigh distribution function.}
\end{figure}

To further investigate the validity of the model, we also performed the dispersion measurement for the band above the mini-gap in Fig. 2(b), for the CROW with $N$ = 400. The experimental data (in the wavelength range shorter than 1540 nm in Fig. 4(b)) agreed well with the numerical result. Hence, we could confirm that the CROW passband was formed based on the mechanism described above.

We observed spectral ripples for both the transmission spectra and the group delay spectra. We considered two possible reasons for these ripples. The first is a Fabry-Perot oscillation caused by an impedance mismatch at the connections between CROW and the access waveguides, due their large $n_{\rm g}$ difference \cite{4}. The second is the effect of cavity wavelength fluctuations $\sigma_{\lambda}$ caused by a fabrication error. The fabrication disorder could degrade the transmission property when $\sigma_{\lambda}$ is comparable or larger than the width of the passband \cite{3}. In our case $\sigma_{\lambda}$ was 1 nm (in a standard deviation), which is smaller than the CROW bandwidth. However, it is important to characterize the uniformity of the cavity structures, which will be valuable information for larger-scale fabrication.

To show the extent of the uniformity, we performed a statistical analysis of the normalized intensity transmittance $\hat{I} = I/\langle I \rangle$. Here $I$ is the experimentally measured transmittance shown in Fig. 3, and $\langle I \rangle$ is the mean transmittance evaluated for each CROW in the spectral region indicated by the arrows in Fig. 3. Then, we calculated the probability distribution function (PDF) of the normalized transmittance $P (\hat{I})$, which is shown as a semi-log plot in Fig. 5.

We see that the PDF exhibited the negative-exponential decay. The behavior agrees well with the Rayleigh distribution $P (\hat{I}) = \exp{(- \hat{I})}$ shown by the dashed line. This suggests that the light transport in the samples is in a ballistic (non-localized) regime, where the fabrication fluctuation of the cavity structures is less relevant than in the localized regime \cite{14, 20}. On the other hand, a strongly disordered structure is known to exhibit a long-tailed log-normal PDF. This suggests the diffusive or localized transport of photons \cite{20,21,22,23,24} due to a structural fluctuation. Our result in Fig. 5 differs greatly from such a distribution. It is noteworthy that the CROW exhibited a non-localized signature up to 400 cavities in a bandwidth of 3.5 nm (0.44 THz). The ballistic transport property has already been confirmed in a CROW formed by 235 Si microring resonators \cite{14}. This time we reveal that a PC-based CROW can also exhibit ballistic transport in the large-scale regime with the present fabrication accuracy, in spite of the fine structures (nanometer-scale displacements of the holes) required for our PC cavities. It is also noteworthy that we could confirm uniformity in the slowest regime ($n_{\rm g} \sim$ 40) ever reported.

In summary, we have investigated the dispersion and light-transport property of a large-scale CROW consisting of high-$Q$ PC nanocavities. We calculated the band structure using the 3-D vector PWE method and explained the formation of the passband in terms of the photonic band picture. The $n_{\rm g}$ spectrum that was experimentally obtained using direct time-domain measurement agreed well with the numerical calculations regardless of the cavity number. A statistical analysis indicated the non-localized light transport property of our CROW up to the 400 cavity chain. Further improvement of the fabrication process would make it possible to implement CROWs with more than a thousand resonators.

We are grateful to Prof. Yasuhiro Tokura and Prof. Shayan Mookherjea for fruitful discussions regarding the light transport.


\end{document}